\documentclass[10pt,twocolumn,letterpaper]{article}

\usepackage{iccv}

\newcommand{\method}{MS-Sync}
\newcommand{\nbenchmarks}{four}

\usepackage{times}
\usepackage{epsfig}

\usepackage{booktabs}
\usepackage{graphicx}%
\usepackage{multirow}%
\usepackage{amsmath,amssymb,amsfonts}%
\usepackage{amsthm}%
\usepackage{mathrsfs}%
\usepackage[title]{appendix}%
\usepackage{xcolor}%
\usepackage{textcomp}%
\usepackage{manyfoot}%
\usepackage{booktabs}%
\usepackage{algorithm}%
\usepackage{algorithmicx}%
\usepackage{algpseudocode}%
\usepackage{listings}%
\usepackage{mathtools}
\usepackage{array}
\newcolumntype{P}[1]{>{\centering\arraybackslash}p{#1}}
\usepackage{eccvabbrv}
\usepackage[T1]{fontenc}

\usepackage[breaklinks=true,bookmarks=false]{hyperref}
\usepackage{cleveref}

\iccvfinalcopy
\begin{document}

\title{A Comprehensive Multi-scale Approach for Speech and Dynamics Synchrony in Talking Head Generation}

\author{
Louis Airale \\
Univ. Grenoble Alpes, CNRS\\
Grenoble INP, LIG \\
38000 Grenoble, France \\
\and
Dominique Vaufreydaz \\
Univ. Grenoble Alpes, CNRS\\
Grenoble INP, LIG \\
38000 Grenoble, France \\
\and
Xavier Alameda-Pineda \\
Univ. Grenoble Alpes, Inria, CNRS\\
Grenoble INP, LJK \\
38000 Grenoble, France \\
}

\maketitle
\ificcvfinal\thispagestyle{empty}\fi

\begin{abstract}
Animating still face images with deep generative models using a speech input signal is an active research topic and has seen important recent progress.
However, much of the effort has been put into lip syncing and rendering quality while the generation of natural head motion, let alone the audio-visual correlation between head motion and speech, has often been neglected.
In this work, we propose a multi-scale audio-visual synchrony loss and a multi-scale autoregressive GAN to better handle short and long-term correlation between speech and the dynamics of the head and lips.
In particular, we train a stack of syncer models on multimodal input pyramids and use these models as guidance in a multi-scale generator network to produce audio-aligned motion unfolding over diverse time scales.
Both the pyramid of audio-visual syncers and the generative models are trained in a low-dimensional space that fully preserves dynamics cues.
The experiments show significant improvements over the state-of-the-art in head motion dynamics quality and especially in multi-scale audio-visual synchrony on a collection of benchmark datasets \footnote{Code and demo available on github page: \href{https://github.com/LouisBearing/HMo-audio}{\texttt{https://github.com/LouisBearing/HMo-audio}}.}
\end{abstract}



\begin{figure*}
  \centering
  \includegraphics[width=0.85\linewidth]{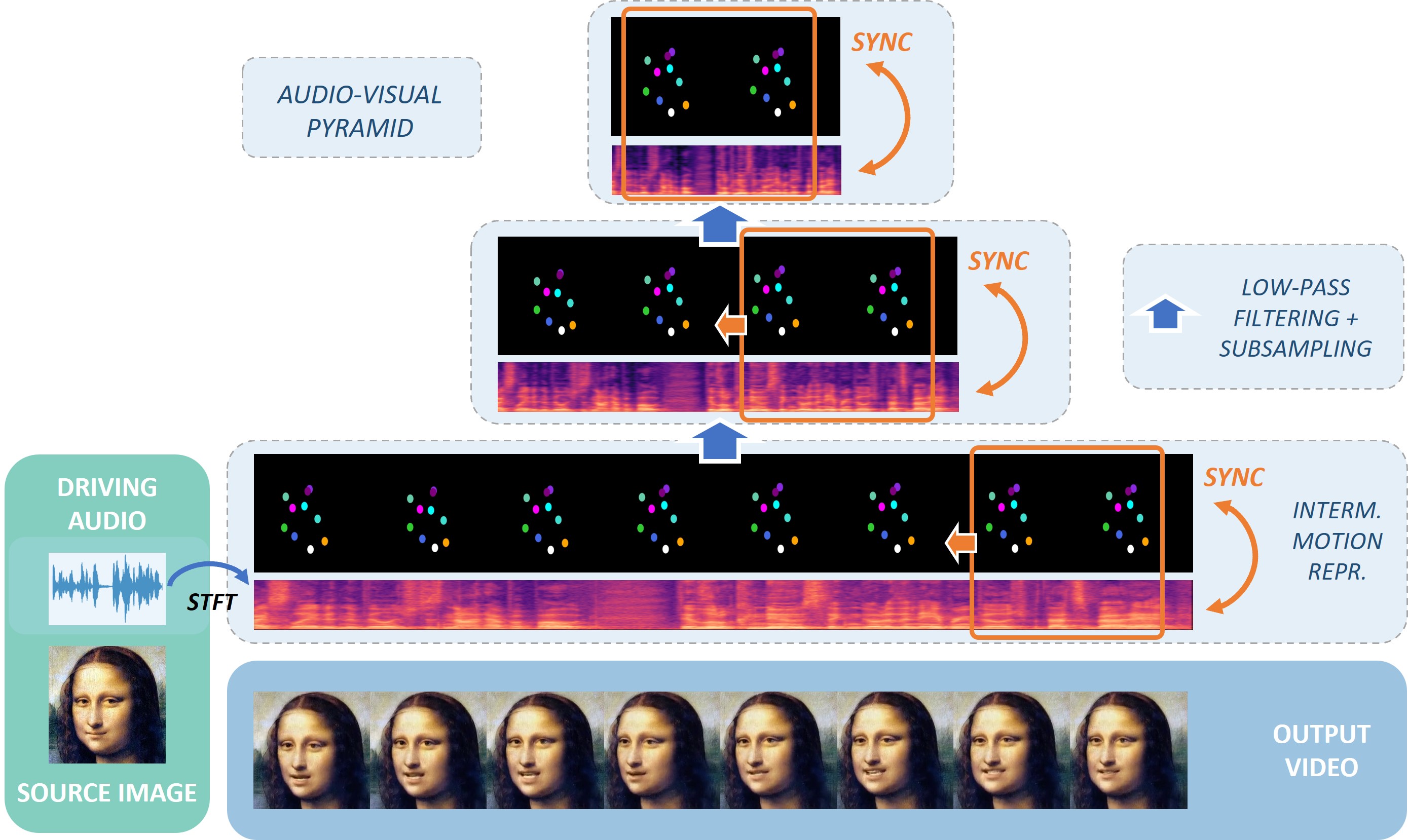}
  \caption{Given a source image and a driving audio signal, our model generates a talking head video sequence correlated with the input speech over multiple time scales, ensuring accurate lips and head dynamics.
  }
  \label{fig:snippet}
\end{figure*}

\section{Introduction}
\label{sec:intro}

The task of talking face generation, which aims to animate still images from a conditioning audio signal, has received considerable attention in the previous years.
The advent of potent reenactment systems, as in Siarohin~\etal~\cite{siarohin2019first} or Wang~\etal~\cite{wang2021one}, and powerful loss functions allowing for a finer correlation between the generated lip motion and the audio input~\cite{chung2017out} have paved the way for a new state of the art.
In both tasks of talking head generation and face reenactment, where lip and head motion are given as a driving video sequence, it is customary to represent face dynamics in a low dimensional space~\cite{greenwood2018joint,chen2019hierarchical,zhou2020makelttalk,zhang2021flow,zakharov2020fast,ha2020marionette,wang2021one,zhao2021sparse,meshry2021learned}.
For this reason recent breakthroughs in face reenactment have also benefited the talking head synthesis task: the above approach assumes that image texture and face dynamics can be processed independently, and that all necessary cues to handle the dynamics fit on a low dimensional manifold.
It is then a reliable strategy to treat audio-conditioned talking face synthesis as a two-step procedure, where the audio-correlated dynamics are first generated in the intermediate space of an off-the-shelf face reenactment model, which is later used to reconstruct photorealistic video samples~\cite{Wang2021Audio2HeadAO,wang2022one,ji2022eamm}.
This allows to focus on improving the audio-visual (AV) correlation between the input speech signal and the produced face and lips movements in a much sparser space than that of real-world images.

Nevertheless, synthesizing natural-looking head and lip motion sequences adequately correlated with an input audio signal remains a challenging task.
In particular, although it has long been known that speech and head motion are tightly associated~\cite{yehia2002linking} (see also Section \ref{rigid_av_correl}), only recently has this relation attracted the attention of the computer vision community.
A likely reason for the difficulty of producing realistic head motion is the lack of an established adequate loss function. 
So far, the most successful strategy to produce synchronized lip movements has relied on the maximization of the cross-modal correlation between short audio and output motion clips, measured by a pre-trained model~\cite{chung2017out,zhu2018arbitrary,prajwal2020lip,park2022synctalkface,yin2022styleheat}.
This fails, however, to account for lower frequency motion as that of the head which remains quasi-static over the short duration considered, typically of the order of a few hundreds of milliseconds.
Surprisingly, there was no attempt to generalize this approach beyond lip synchronization.
Neither has possible multi-scale audio-visual correlation been explored in the talking face generation literature.
Head motion is often produced through the use of a separate sub-network trained to match the dynamics of a ground truth sequence, which in practice decouples the animation of head and lips.

We argue that to account for motion that unfolds over longer durations such as the head rhythm, a dedicated loss enforcing the synchrony of AV segments of various lengths is needed.
We propose to implement this loss using a \textit{pyramid of syncers}, replacing the lip-sync expert of Prajwal \etal~\cite{prajwal2020lip} with a stack of syncer models evaluating the correlation between the audio input and the dynamics of the whole face over different time scales.
How to achieve this however is not trivial, as simply increasing the length of the video segments does not guarantee that the expert model will focus on lower frequency motion.
Applying a low-pass filtering on the video input, on the other end, would produce blurry and unusable results.
Conversely, one may readily compute a multi-scale representation of motion by successively filtering a sequence of facial points coordinates.
We take advantage of this observation in two different ways: 1) by using smoothed facial coordinate sequences as input to the pyramid of syncers, 2) by producing facial coordinates as output from the generative model.

Concretely, we propose to construct Gaussian pyramids of head dynamics and audio on two sets of inputs: first, on paired samples from an audio-visual dataset for the training of the syncers, then on the generated dynamics and their corresponding driving audio signal during generative model training.
The resulting \textit{multi-scale audio-visual synchrony loss} hence represents a powerful means to enforce the correlation of both rigid, low-frequency and non-rigid, high-frequency generated motion with the input speech.
Here, another advantage of operating on a low dimensional motion manifold is that the loss can be applied directly on the model's output without propagating through the visual rendering network, significant accelerating training.
In addition, the multi-scale AV synchrony loss allows to produce head and lip movements with a single network, resulting in overall lighter architecture and training procedure compared to previous works that use an \textit{ad hoc} network to model head movement.

To exploit the gradients from the multi-scale syncers, we build a hierarchical generative model, using a Feature Pyramid Network (FPN)~\cite{lin2017feature} backbone.
We use an autoregressive model for its flexibility to handle sequences of arbitrary length, and complement the AV synchrony loss with a window-based multi-scale discriminator architecture that proved to perform well on the generation of facial landmarks~\cite{airale2022autoregressive}.
The resulting method, hereafter labeled \method{} (for Multi-Scale Synchrony), produces head dynamics in a low dimensional motion space, namely that of 2d unsupervised keypoints.
Videos are then reconstructed using an off-the-shelf, frozen reenactment model~\cite{siarohin2019first}.
We would however like to point out that the multi-scale AV synchrony loss is a standalone contribution that can fuse in diverse generative model architecture, making it a versatile tool for audio-driven talking head generation.



The main contributions of the present work are summarized below.
\begin{itemize}
    \item We train audio-visual syncer networks on rigid head motion, measuring for the first time, to the best of our knowledge, the correlation between head motion and speech,
    \item A multi-scale audio-visual synchrony loss to enable the generation of diverse audio-correlated facial dynamics,
    \item A multi-scale autoregressive GAN~\cite{goodfellow2014generative} framework, labelled \method{}, effective at producing speech-synchronized head and lips motion as shown in extensive experiments on \nbenchmarks{} benchmark datasets.
\end{itemize}

\section{Related Work}

\subsection{Talking Head Generation}

The task of talking head generation consists of animating a source image or an initial face using a driving audio signal, and is especially difficult as it supposes to produce audio-synced head and lip motion while preserving high visual quality results.
Hence trade-offs usually need to be done on one aspect or the other, or alternatively on the inference duration or the generalizability to unseen identities, and talking head generation methods therefore come in many different flavors.

An important part of the literature consists of identity-dependent methods, that typically require fine-tuning on a target identity at inference~\cite{yi2020audio,chen2020talking,zhang2021facial}.
Those comprise pioneering research works~\cite{suwajanakorn2017synthesizing,karras2017audio}, but also many recent approaches based on the NeRF framework~\cite{mildenhall2021nerf} that although providing outputs of compelling realism, require fine-tuning on video clips of the source subject~\cite{guo2021ad,shen2022learning,liu2022semantic,ye2023geneface,li2023efficient}.
Another class of high visual quality methods is based on diffusion models~\cite{sohl2015deep,ho2020denoising,du2023dae}, that typically trade the generation of head pose for that of visual sharpness.
Indeed head pose is either provided as a driving signal~\cite{shen2023difftalk}, or only weakly controlled with a mean-square-error (MSE) loss in the visual domain, giving less accurate audio correlation~\cite{stypulkowski2024diffused,yu2023talking}.

A second, overlapping part of the literature focuses on the lip-syncing accuracy: head pose is either extracted from a driving video sequence, possibly with the lip region masked out, or it is simply omitted~\cite{eskimez2018generating,song2018talking,zhu2018arbitrary,chen2019hierarchical,zhou2019talking,vougioukas2020realistic,doukas2021headgan,ji2022eamm,sinha2022emotion,fan2022faceformer,liang2022talkingflow,xing2023codetalker,zhong2023identity,tan2023emmn,wang2023lipformer,huang2023parametric,xu2023high}.
These methods provide strong audio-correlation baselines, and a typical strategy is to use the output of a frozen lip-sync model to improve the synchrony between output lips and input speech signal~\cite{zhou2021pose,liang2022expressive,wang2022progressive,yin2022styleheat,park2022synctalkface,wang2023seeing}.

Finally, close to the proposed approach are one-shot methods that reenact single source images of unseen identities in real time, with an explicit treatment of head motion~\cite{zhou2020makelttalk,Wang2021Audio2HeadAO,zhang2021flow,liu2023moda}.
Although successful attempts have been made to leverage pre-trained syncer models for precise lip-syncing~\cite{wang2022one,zhang2023sadtalker}, head pose is still usually learned through the minimization of a MSE loss that fails to explicitly account for the correlation between speech and rigid head motion.
Besides, we argue that the duration of the audio segments commonly used for the synchronization of the lips is insufficient to properly align lower-frequency movements like that of the head with the driving audio signal, advocating for novel approaches.

\subsection{Video-Driven Face Reenactment}

The task of animating a source human face with a neural network can also be fully guided by a driving sequence of a target identity that provides the supervision for head and lip motion~\cite{zakharov2019few,ha2020marionette,zhao2021sparse,meshry2021learned}.
Compelling results have been achieved over the years to bridge the identity gap between source and target images and hallucinate unseen poses of the source identity~\cite{siarohin2019first,guo2021ad,ren2021pirenderer}.
These works rely on low dimensional representations, e.g. facial landmarks~\cite{zhao2021sparse,meshry2021learned,zakharov2020fast} or learned keypoints~\cite{siarohin2019first,wang2021one} to measure the deformation from a given target image to the source image, which is later used to warp or normalize the style of the source identity image.
Following previous works~\cite{Wang2021Audio2HeadAO,wang2022one}, we train a generative model to predict the speech-conditioned driving motion sequence, and use the reenactment model of Siarohin~\etal~\cite{siarohin2019first} to reconstruct the output videos.

\subsection{Multi-scale Data Processing}

Learning on representations of the input data over multiple time or spatial scales has become the standard in computer vision tasks such as object detection or semantic segmentation where objects of the same class can have different sizes~\cite{lin2017feature,tan2020efficientdet}. 
In the generative models literature, multi-scale approaches may either be implemented in the discriminator network of GANs as a way to improve multi-scale faithfulness of generated data~\cite{wang2018high,lin2018human,kumar2019melgan} but also in the generative model itself~\cite{denton2015deep,karras2019style}.
Although this was not explored so far in talking head generation, multi-scale feature hierarchies can be readily computed to align speech and dynamics of various motion frequencies. 

\begin{figure*}
\includegraphics[width=1.0\linewidth]{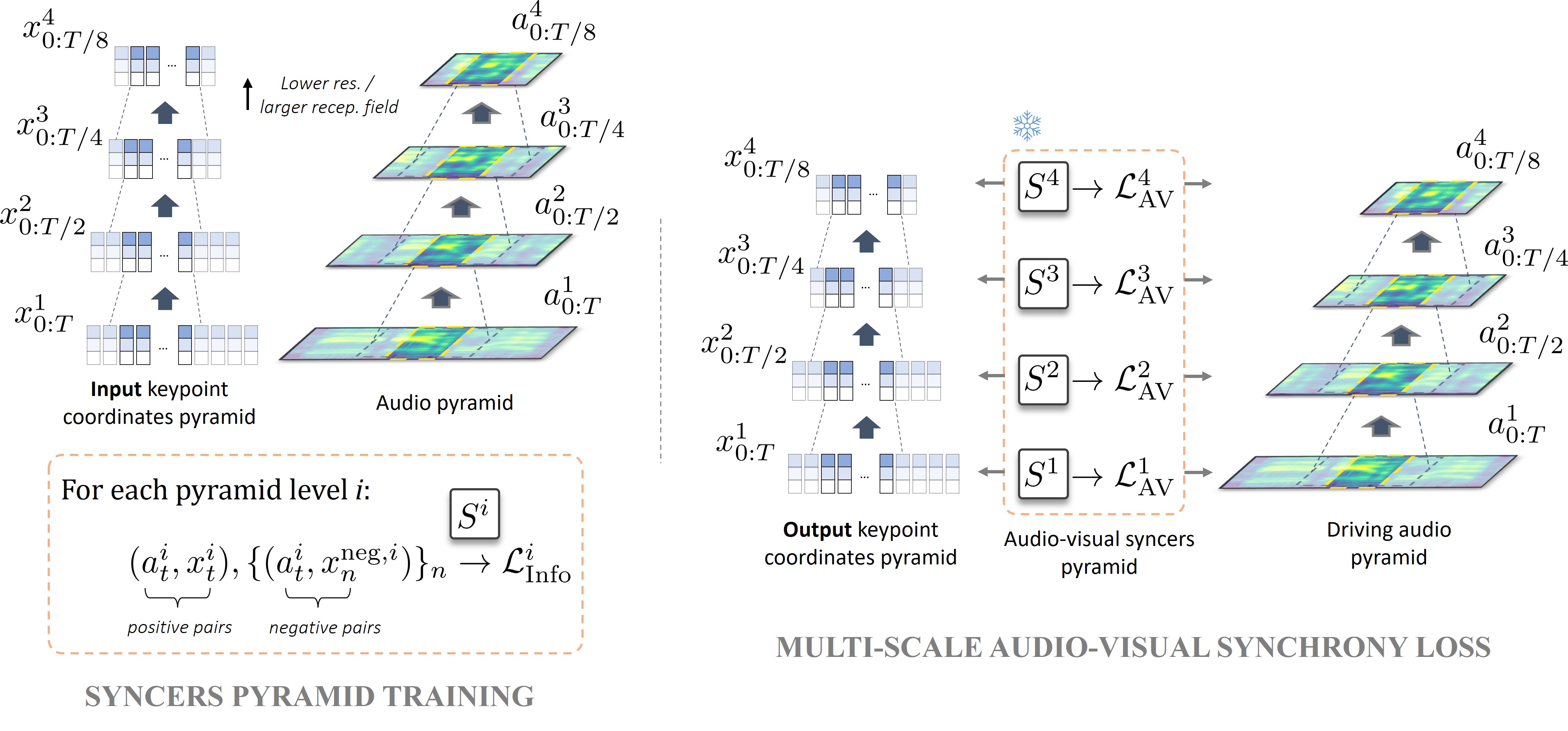}
\centering
\caption{\textbf{Left}. A stack of syncer networks $S^i$ are trained on multi-scale positive and negative multimodal pairs using contrastive losses. \textbf{Right}. The syncer models are frozen and used to compute the multi-scale audio-visual synchrony loss of the generative model.}
\label{fig:ms_av_loss}
\end{figure*}

\section{Method}
\label{sec:method}

This section introduces the problem formulation and associated notations.
We follow the conventions from Siarohin~\etal~\cite{siarohin2019first} and represent the dynamics using a set of $10$ 2d keypoints coordinates, together with two-by-two Jacobian matrices that provide finer details of the spatial deformation around the keypoints.
In the following the overall inputs are simply referred to as keypoints, of total dimension $60$.
Given a set of initial keypoints coordinates $x_0 \in \mathbb{R}^{60}$ and a conditioning audio signal $a_{0:T} = (a_0, \ldots, a_T) \in \mathbb{R}^{d \times T}$ (here $d=26$) over $T$ time steps, we aim to produce a sequence of keypoint positions $x_{1:T}$ such that the joint distributions over generated and data samples match:
\begin{equation}
p_{g}(x_{0:T},a_{0:T}) = p_{\text{data}}(x_{0:T},a_{0:T}), \quad \forall x_{0:T}, a_{0:T}.
\end{equation}

The procedure to tackle this problem is as follows.
The multi-scale AV synchrony loss, which is the major contribution of this work, is first introduced in~\cref{sec:ms_av_loss}.
Then a multi-scale generator architecture able to exploit appropriately the devised multi-scale AV loss is developed in~\cref{sec:ms_gen}.
Finally the overall training procedure is detailed in~\cref{sec:overall_archi}.

\subsection{Multi-scale Audio-Visual Synchrony Loss}
\label{sec:ms_av_loss}

The most prominent procedure to align dynamics with speech input relies on the optimization of a correlation score computed on short audio-visual segments of the generated sequence using a pre-trained AV syncer network~\cite{prajwal2020lip}.
Several contrastive loss formulations are possible to train the syncer network, that suppose the maximization of the agreement between in-sync AV segments or positive pairs $(a_t, x_t)$ versus that of out-of-sync or negative pairs.
One particularly interesting formulation is the Info Noise Contrastive Estimation loss, which maximizes the mutual information between its two input modalities~\cite{oord2018representation}.
Given a set $X = (a_t, x_t, x_1^{neg}, \ldots, x_N^{neg})$ containing a positive pair and $N$ negative position segments, this loss writes:
\begin{equation}
\label{eq:infonce}
    \mathcal{L}_{\text{InfoNCE}} = - \mathbb E_X \frac{e^{S(a_t, x_t)}}{e^{S(a_t, x_t)} + \sum_{n=1}^N e^{S(a_t, x_n^{neg})}},
\end{equation}
with $S$ the syncer model score function, which is classically implemented hereafter as the cosine similarity of the outputs from an audio and a position embeddings $e_a$ and $e_x$:
\begin{equation}
    S(a_t, x_t) = \frac{e_a(a_t)^\top e_x(x_t)}{\lVert e_a(a_t)\rVert.\lVert e_x(x_t)\rVert}.
\end{equation}
Following the usual practice, $a_t$ and $x_t$ are respectively taken as the MFCC spectrogram and position segment of a 200 ms window centered on time step $t$.
Negative pairs can be misaligned audio and position segments from the same audio-visual sequence and therefore constitute hard negatives, or can alternatively be segments from different samples, \eg when an insufficient number $N$ of hard negatives can be sampled.

Once trained, the weights of $e_a$ and $e_x$ are frozen and the following term is added
to the loss function of the generative model:
\begin{equation}
\label{eq:av_loss}
    \mathcal{L}_{\text{AV}} = - \mathbb E_t S(a_t, x_t),
\end{equation}
where $a_t$ is now part of the conditioning signal and \textit{$x_t$ is output by the model}.

The above procedure is insufficient when one needs to discover AV correlations over different time scales. 
One solution consists in building multi-scale representations of the audio-visual inputs and training one syncer network $S^i$ for each level $i$ in the resulting pyramid.
Here the use of keypoints instead of the raw video is crucial for two reasons: they can be easily downscaled by successive low-pass filtering operations, and it avoids propagating through the visual reenactment model, substantially saving computation.
The training process of the pyramid of syncers is represented in~\cref{fig:ms_av_loss} (left).
Specifically, audio and keypoint coordinates pyramids $\{a_{0:T/2^{i-1}}^i\}_i$ and $\{x_{0:T/2^{i-1}}^i\}_i$ are constructed by successive passes through an average pooling operator that blurs and downscales its input by a factor 2, e.g. for positions:
\begin{equation}
    x_t^i = \frac{1}{2k + 1} \sum_{\tau=-k}^k x_{2t + \tau}^{i-1}
\end{equation}
where we choose $k=3$.
The objective is to progressively blur out the highest frequency motion when moving upward in the pyramid, forcing the top level syncers to exploit better the rhythm of the head motion.
A total of four syncer networks are trained on the input pyramid following~(\ref{eq:infonce}), input segment duration ranging from the standard 200 ms on the bottom level to 1600 ms at the coarsest scale.

After the training of the pyramid of syncers, all networks $S^1$ to $S^4$ are frozen and used to compute the multi-scale audio-visual synchrony loss.
The principle of this loss is presented in~\cref{fig:ms_av_loss}.
Similar to the input pyramids used to train the syncer networks, we construct a multi-scale representation of the input speech $a_{0:T}$ and the generated keypoints positions $x_{0:T}$.
Then for each hierarchy level $i$ one loss term $\mathcal{L}_{\text{AV}}^i$ is computed according to~(\ref{eq:av_loss}) using pre-trained syncer $S^i$.
Those terms are then summed to give the overall multi-scale AV synchrony loss $\mathcal{L}_{\text{AV}}^{\text{MS}}$:
\begin{align}
&\mathcal{L}_{\text{AV}}^{\text{MS}} = \sum_i \mathcal{L}_{\text{AV}}^i(x_{0:T/2^{i-1}}^i, a_{0:T/2^{i-1}}^i), \\
&\text{with} \ \mathcal{L}_{\text{AV}}^i(x_{0:T/2^{i-1}}^i, a_{0:T/2^{i-1}}^i) = - \mathbb{E}_t \bigl[ S^i(a_t^i, x_t^i) \bigr]
\end{align}
To better exploit the effects of this loss, we propose a multi-scale autoregressive generator network which is described in the following section.

\begin{figure*}
\includegraphics[width=0.9\linewidth]{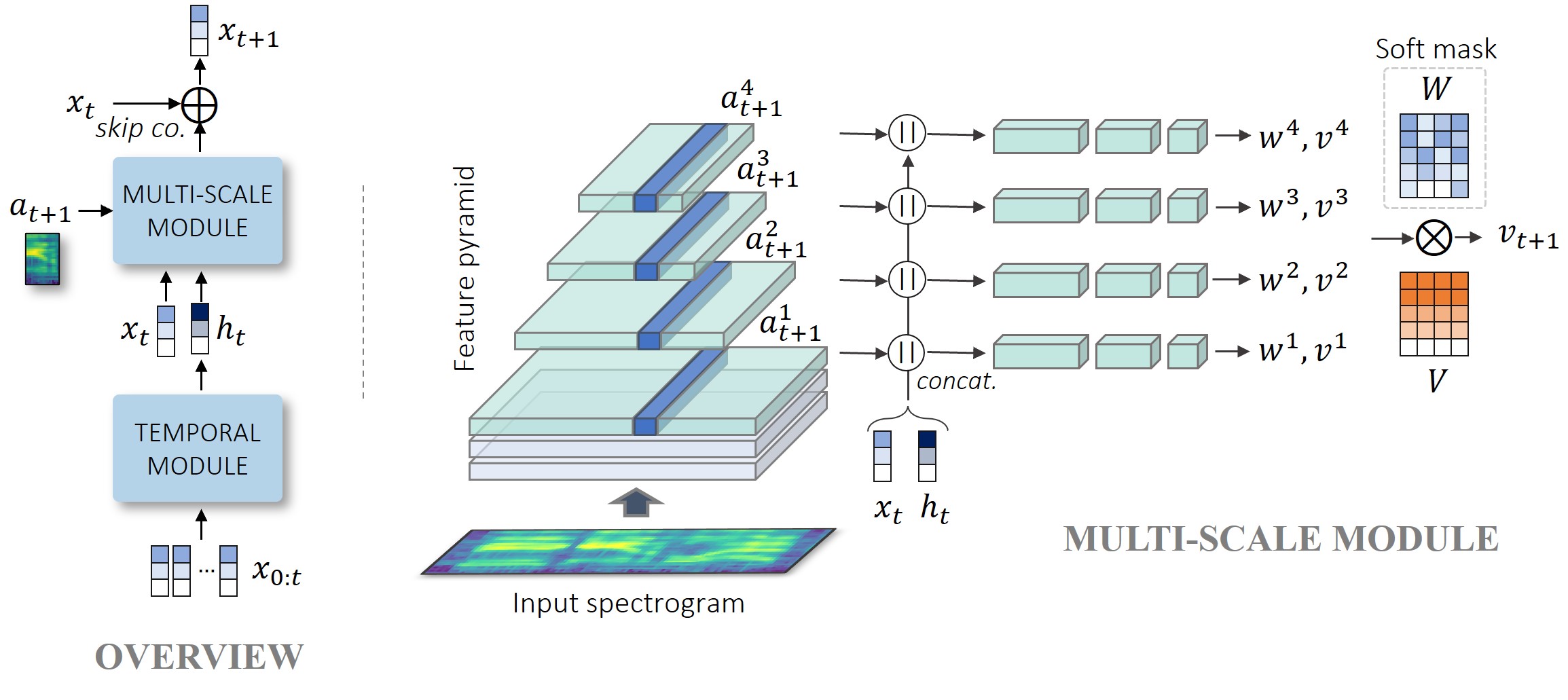}
\centering
\caption{\textbf{Left}. Our network is composed of an autoregressive Transformer temporal module and a convolutional multi-scale module. \textbf{Right}. Details of the multi-scale module.}
\label{fig:g_archi}
\end{figure*}

\subsection{Multi-scale Autoregressive Generator}
\label{sec:ms_gen}

Through the multi-scale synchrony loss, the generator receives gradients that push it to produce audio-synced keypoint positions over multiple time scales.
In this section, we describe the architecture of the generator network, which is itself implemented with a multi-scale structure to allow distinct loss terms to act preferentially on different layers of the network.
The overall architecture is depicted in~\cref{fig:g_archi}.

A residual autoregressive formulation is employed for the generative model, such that given keypoints positions $x_{0:t}$ up to time step $t$ and the audio input $a_{t+1}$ for the next time step, the generator $G$ produces instantaneous velocities $v_{t+1}$:
\begin{align}
    v_{t+1} &= G(x_{0:t}, a_{t+1}), \\
    x_{t+1} &= x_t + v_{t+1}.
\end{align}

As depicted in~\cref{fig:g_archi}, $G$ contains a temporal module that operates on a sequence of keypoint positions, and a multi-scale module that takes the output of the temporal module $h_t$, the positions $x_t$ and audio $a_{t+1}$ as input to produce $v_{t+1}$.
A Transformer encoder~\cite{vaswani2017attention} is used for the temporal module, and the multi-scale module is implemented as the bottom-up path of a Feature Pyramid Network~\cite{lin2017feature}.
Namely, the input spectrogram is processed by several downsampling convolutional layers, producing feature maps $a_{0:T}^1$ to $a_{0:T/2^3}^4$ of the same resolution as those used to compute the pyramids for the audio-visual synchrony loss.
Feature maps 2 to 4 are later interpolated back to the length $T$ of the finest map, such that one vector $a_{t+1}^i$ can be extracted from each pyramid level $i$ to produce the next step velocity.
Concretely, each vector $a_{t+1}^i$ is concatenated with $x_t$ and $h_t$ and is processed by an independent fully connected branch, the rationale being that processing each input resolution separately would allow the model to produce different motion frequencies.

The outputs of the four branches of the multi-scale generator are merged using a learnable soft spatial mask.
Each branch $i$ outputs a velocity vector $v^i \in \mathbb R^{10\times6}$ (recall that there are both 2d coordinates and 4d Jacobian matrices, and time index is omitted for the sake of clarity) and a mask vector $w^i \in \mathbb R^{10\times6}$ such that $w^i_{j,k} = w^i_{j,l}$ for all $j$, $k$ and $l$ (or one mask value for all 6 coordinates of a given keypoint), responsible for enhancing or weakening the contribution of each keypoint on the given branch.
This is because facial regions are expected to play different roles depending on the scale: the finest resolution branch might emphasize lip keypoints, while at the coarsest scale, more weight may be put on rigid head motion.
The output of the multi-scale module finally writes:
\begin{equation}
    v_{t+1} = \sum_{i=1}^4 \Bigl(\frac{e^{w^i}}{\sum_j e^{w^j}}\Bigl) v^i
\end{equation}








\subsection{Overall Architecture and Training}
\label{sec:overall_archi}

The audio-visual synchrony loss $\mathcal{L}_{\text{AV}}^{\text{MS}}$ is complemented with two discriminator networks to improve the static and dynamic quality of the generated keypoints.
One frame discriminator $D_f$ computes the realism of static keypoints, and a window-based multi-scale network $D_s$ computes that of keypoint sequences~\cite{airale2022autoregressive}.
Adversarial losses are implemented with the geometric GAN formulation of Lim \& Ye~\cite{lim2017geometric}.
Namely, given the generated and ground truth keypoint position distributions $p_g$ and $p_{\text{data}}$, the generator losses write:
\begin{align}
    \mathcal{L}_{G_f} &= -\mathbb E_{x_{0:T} \sim p_g} \left[ \frac{1}{T} \sum_{t \geq 1} D_f(x_t) \right], \\
    \mathcal{L}_{G_s} &= -\mathbb E_{x_{0:T} \sim p_g} \left[ D_s(x_{0:T}) \right],
\end{align}
as for the generic discriminator loss:
\begin{multline}
    \mathcal{L}_{D_*} = \mathbb E_{x \sim p_g} \left[\max(0, 1 + D_*(x)) \right] \\ + \mathbb E_{x \sim p_{\text{data}}} \left[\max(0, 1 - D_*(x)) \right],
\end{multline}
where $D_*$ is replaced respectively by $D_f$ and $D_s$ .
A $L_2$ reconstruction loss to the ground truth keypoints, $\mathcal{L}_{\text{rec}}$, is finally added to the loss function:
\begin{equation}
    \mathcal{L}_{\text{rec}} = \mathbb E_{(x_{0:T}, a_{0:T}) \sim p_{\text{data}}} \lVert x_{0:T} - G(x_0, a_{0:T}) \rVert^2
\end{equation}
The balance between loss terms is achieved by the use of weighting factors $\lambda_{\text{av}}$, $\lambda_{\text{adv}}$ and $\lambda_{\text{rec}}$ (with $\lambda_{\text{av}} = 8$, $\lambda_{\text{adv}} = 0.1$ and $\lambda_{\text{rec}} = 1$ providing the best results), such that the overall training consists in minimizing alternatively the two following terms:
\begin{align}
    &\mathcal{L}_D = \mathcal{L}_{D_f} + \mathcal{L}_{D_s} \\
    &\mathcal{L} = \lambda_{\text{av}} \mathcal{L}_{\text{AV}}^{\text{MS}} + \lambda_{\text{adv}} (\mathcal{L}_{G_f} + \mathcal{L}_{G_s}) + \lambda_{\text{rec}} \mathcal{L}_{\text{rec}}.
\end{align}

\section{Experiments}

We conduct benchmark evaluations to measure the proficiency of our method on visual audio-visual synchrony, landmark-domain multi-scale audio-visual synchrony and output visual quality.
We also carry an ablation study to investigate the contribution of the different terms of the loss function, including the multi-scale AV loss.

\subsection{Experimental protocol}

\paragraph{Datasets.}
Experiments are conducted on the VoxCeleb2 dataset~\cite{Chung18b} with two different preprocessings.
First, we use the standard test set of VoxCeleb2, hereafter VoxCeleb2~(I), containing $\sim$2000 short audio-visual clips centered on subject faces.
Second, following the preprocessing strategy of Siarohin~\etal~\cite{siarohin2019first}, subsets of respectively $\sim$18k and 500 short video clips from the original VoxCeleb2 train and test sets are generated, the former for the training of our model, and the latter for the evaluation.
The interest of this preprocessing is that it keeps the reference frames fixed, thus preserving head motion.
In the following section, this second dataset is referred to as VoxCeleb2 (II).
Evaluations are also conducted on the HDTF dataset~\cite{zhang2021flow}, which contains $\sim$400 long duration frontal-view videos from political addresses, where head motion is also preserved but span a rather narrow distribution.
Last, we use LRS2~\cite{Afouras18c}, which is preprocessed similarly to VoxCeleb2~(I), to measure the audio-visual synchrony in the visual domain.

\paragraph{Benchmark Models.}

We compare our method, \method{}, with other one-shot talking head generation models.
Wav2Lip~\cite{prajwal2020lip} uses a pre-trained lip syncer to 
learn the AV synchrony, and achieved state-of-the-art performances on the visual dubbing task.
However, it only reenacts the lip region and therefore does not produce any head motion. IP\_LAP~\cite{zhong2023identity} does not produce head motion either, contrary to PC-AVS~\cite{zhou2021pose} and EAMM~\cite{ji2022eamm}, albeit being rather limited.
On the other hand, MakeItTalk~\cite{zhou2020makelttalk}, Audio2Head~\cite{Wang2021Audio2HeadAO} and its follow-up work~\cite{wang2022one}, and SadTalker~\cite{zhang2023sadtalker} output head poses.
Noticeably, the two methods presented in Wang~\etal\cite{Wang2021Audio2HeadAO} and Wang~\etal~\cite{wang2022one} rely on the same facial keypoints and reenactment model as in the present work, although here the reenactment model is not fine-tuned.

\paragraph{Training Details.}

The temporal module introduced in~\cref{sec:method} is implemented using 4 self-attention layers with 8 heads each.
Both audio and keypoints are encoded as 512-dimensional vectors, and both syncers and the generative models were trained on VoxCeleb2~(II).
The frame discriminator $D_f$ is a multi-layer perceptron, and the window-based multi-scale $D_s$ follows the LSTM implementation of Airale~\etal~\cite{airale2022autoregressive}.
Syncers training lasts from around one day for the finest time scale to a few hours for the coarsest model.
We rely on hard negative mining for the first three time scales with $N=12$, and use negative pairs from different audio and dynamics samples at the coarsest scale where the shortened sequences reduce the number of available hard negative pairs, and set $N=48$ for this last model.
Syncers are then frozen and used for the training of the generative model, which is trained to predict sequences of 40 frames for 70k iterations (about 500 epochs) using Adam optimizers with $\beta_1=0$ and $\beta_2=0.999$ and learning rates of $2\times 10^{-5}$ and $1\times 10^{-5}$ respectively for the generator and the discriminator, after which a decay factor of 0.1 is applied on the learning rates for 5k additional iterations.
All audio inputs are sampled at 16~kHz, and we use a window size of 400 and hop size of 160 to generate the 26-dimensional MFCC spectrograms.

\subsection{Image-Domain AV Synchrony}

\paragraph{Protocol.}
The first benchmark follows the customary evaluation of the audio-visual synchrony in the visual domain using several standard metrics, on both VoxCeleb2~(I) and LRS2.
To cope with the imbalanced duration of VoxCeleb2 videos and keep computation time manageable, we work with the first 40 frames of each test clip, while we use the whole LRS2 test set, which contains shorter videos (41 frames in average, ranging from 15 to 145 frames).
We use the absolute offset $|\text{AV-Off}|$ and the confidence score AV-Conf output by SyncNET~\cite{chung2017out}, and also extract the facial landmarks from the output videos~\cite{bulat2017far} to compute the frontalized landmark distance (LMD) from the predicted mouth region to the ground truth landmarks, by rotating each frame to a canonical pose.
This ensures that all methods are placed on an equal footing, whether they produce head motion or not.
Results are presented in Table~\ref{tab:visual_av_vox}.

\begin{table*}[t]
\caption[]{Image domain AV synchrony. LMD is the frontalized landmark distance, with the face rotated back to a canonical pose.$\dagger$ We rescale PC-AVS by a factor 0.75 to account for cropping. *Wav2Lip is trained to optimize the SyncNet scores, hence the very strong confidence.
\footnotemark}
\label{tab:visual_av_vox}
\begin{center}
\resizebox{1.0\textwidth}{!}{
\begin{tabular}{l c c c | c c c }
\toprule
 \multicolumn{1}{r}{Dataset} & \multicolumn{3}{c}{VoxCeleb2 (I)} & \multicolumn{3}{c}{LRS2} \\
 \cmidrule(lr){2-4} \cmidrule(lr){5-7}
Method & $|\text{AV-Off}| \downarrow$ & $\text{AV-Conf} \uparrow$ & $\text{LMD} \downarrow$ & $|\text{AV-Off}| \downarrow$ & $\text{AV-Conf} \uparrow$ & $\text{LMD} \downarrow$\\
\midrule

Ground truth & $1.89 {\pm 1.92}$ & $6.29 {\pm 1.66}$ & 0.0
& $0.08 {\pm 0.4}$ & $8.36 {\pm 1.62}$ & 0.0\\

\midrule

MakeItTalk~\cite{zhou2020makelttalk}& $5.23 {\pm 4.29}$ & $3.50 {\pm 1.49}$ & $2.80 {\pm 1.10}$
& $8.43 {\pm 6.16}$ & $2.56 {\pm 0.96}$ & $2.80 {\pm 1.07}$ \\

Wav2Lip*~\cite{prajwal2020lip}&$2.86 {\pm 0.34}$ & $\textbf{8.07} {\pm \textbf{1.33}}$ & $2.41 {\pm 0.77}$ 
& $2.79 {\pm 0.54}$ &$\textbf{8.72} {\pm \textbf{1.25}}$ & $2.54 {\pm 0.70}$\\

$\text{PC-AVS}^\dagger$~\cite{zhou2021pose}& $5.18 {\pm 3.31}$ & $3.85 {\pm 1.55}$ & $2.65 {\pm 1.23}$
& $5.48 {\pm 3.65}$ & $4.42 {\pm 1.65}$ & $2.61 {\pm 0.85}$\\

Audio2Head~\cite{Wang2021Audio2HeadAO}& $6.83 {\pm 6.66}$ & $2.66 {\pm 1.38}$ & $3.23 {\pm 1.19}$
& $6.78 {\pm 6.72}$ & $3.18 {\pm 1.43}$ & $3.20 {\pm 1.09}$\\

EAMM~\cite{ji2022eamm}& $9.52 {\pm 5.76}$ & $1.94 {\pm 0.75}$ & $3.30 {\pm 1.24}$
& $9.07 {\pm 5.94}$ & $2.41 {\pm 0.87}$ & $3.50 {\pm 1.18}$\\

Wang~\etal~\cite{wang2022one}& $2.59 {\pm 4.29}$ & $4.12 {\pm 1.72}$ & $2.89 {\pm 1.26}$
& $\underline{2.59} {\pm \underline{4.23}}$ & $4.56 {\pm 1.67}$ & $2.89 {\pm 1.26}$\\

IP\_LAP~\cite{zhong2023identity}& $4.78 {\pm 5.01}$ & $3.15 {\pm 1.40}$ & $\textbf{2.34} {\pm \textbf{0.79}}$
& $4.73 {\pm 5.03}$ & $3.80 {\pm 1.57}$ & $2.41 {\pm 0.71}$\\

SadTalker~\cite{zhang2023sadtalker}& $\underline{2.19} {\pm \underline{3.92}}$ & $4.63 {\pm 1.91}$ & $2.38 {\pm 0.75}$
& $2.72 {\pm 4.60}$ & $5.01 {\pm 1.97}$ & $\textbf{2.28} {\pm \textbf{0.86}}$\\

\midrule

\method{} (Ours) & $\textbf{1.00} \pm \textbf{1.64}$ & $\underline{5.81} {\pm \underline{1.66}}$ & $\underline{2.36} \pm \underline{0.73}$
& $\textbf{1.06} \pm \textbf{2.14}$ & $\underline{5.75} {\pm \underline{1.69}}$ & $\underline{2.34} \pm \underline{0.81}$\\
\bottomrule
\end{tabular}
}
\end{center}
\end{table*}

\begin{table*}[t]
\caption{Landmark domain rigid multi-scale AV synchrony on VoxCeleb2 (II), where test syncers are trained on rigid facial parts: eyes and nose are considered, but not lips and jaws landmarks.}
\label{tab:head_av_vox}
\begin{center}
\resizebox{1.0\textwidth}{!}{
\begin{tabular}{l c c c | c c c}
\toprule
 \multicolumn{1}{l}{Dataset} & \multicolumn{3}{c}{VoxCeleb2 (II)} & \multicolumn{3}{c}{HDTF}\\
 \midrule
 \multicolumn{1}{r}{Time scale} & \multicolumn{1}{c}{200 ms (1)} & \multicolumn{1}{c}{400 ms (2)} & \multicolumn{1}{c}{800 ms (3)} & \multicolumn{1}{c}{200 ms (1)} & \multicolumn{1}{c}{400 ms (2)} & \multicolumn{1}{c}{800 ms (3)}
 \\

 \cmidrule(lr){2-2} \cmidrule(lr){3-3} \cmidrule(lr){4-4} \cmidrule(lr){5-5} \cmidrule(lr){6-6} \cmidrule(lr){7-7}
 \multicolumn{1}{l}{Method} & $|\text{AV-Off}_1| \downarrow$ & $|\text{AV-Off}_2| \downarrow$ & $|\text{AV-Off}_3| \downarrow$ & $|\text{AV-Off}_1| \downarrow$ & $|\text{AV-Off}_2| \downarrow$ & $|\text{AV-Off}_3| \downarrow$ \\
\midrule

Random & $10.59 {\pm 4.18}$ & $10.50 {\pm 4.09}$ & $10.11 {\pm 4.03}$ & $12.14 {\pm 4.01}$ & $10.32 {\pm 4.56}$ & $9.48 {\pm 4.63}$\\

Ground truth & \hphantom{0}$4.57 {\pm 5.32}$ & \hphantom{0}$4.64 {\pm 5.60}$ & \hphantom{0}$7.48 {\pm 5.47}$ & \hphantom{0}$9.43 {\pm 5.53}$ & \hphantom{0}$7.63 {\pm 6.00}$ & \hphantom{0}$7.89 {\pm 5.74}$\\

\midrule

MakeItTalk~\cite{zhou2020makelttalk}& \hphantom{0}$6.91 {\pm 5.95}$ & \hphantom{0}$6.77 {\pm 6.09}$ & \hphantom{0}$8.64 \pm 5.42$ & $14.40 {\pm 1.90}$ & $13.64 \pm 2.62$ & $14.18 \pm 2.05$  \\

Wav2Lip~\cite{prajwal2020lip} &  \hphantom{0}$5.38 {\pm 5.90}$ & \hphantom{0}$6.72 \pm 6.28$ & \hphantom{0}$9.22 \pm 5.52$ & $15.00 {\pm 0.05}$ & $14.88 \pm 0.61$ & $14.87 \pm 0.72$ \\

Audio2Head~\cite{Wang2021Audio2HeadAO} & \hphantom{0}$9.43 {\pm 4.79}$ & \hphantom{0}$8.95 {\pm 5.09}$ & \hphantom{0}$9.61 {\pm 4.91}$ & \hphantom{0}$\underline{9.55} {\pm \underline{5.40}}$ & \hphantom{0}$8.60 {\pm 5.55}$ & \hphantom{0}$7.97 {\pm 5.30}$  \\

EAMM~\cite{ji2022eamm} & \hphantom{0}$\underline{4.98} {\pm \underline{5.71}}$ & \hphantom{0}$\underline{4.71} {\pm \underline{5.87}}$ & \hphantom{0}$\underline{7.30} {\pm \underline{5.80}}$ & $14.09 {\pm 2.39}$ & $13.78 {\pm 2.49}$ & $14.30 {\pm 1.76}$ \\

Wang~\etal~\cite{wang2022one} & \hphantom{0}$8.26 {\pm 5.61}$ & \hphantom{0}$8.58 {\pm 5.58}$ & \hphantom{0}$9.14 {\pm 5.07}$ & $10.68 {\pm 4.42}$ & \hphantom{0}$\underline{8.55} {\pm \underline{4.55}}$ & \hphantom{0}$\underline{7.21} {\pm \underline{4.78}}$ \\

IP\_LAP~\cite{zhong2023identity} & \hphantom{0}$9.66 {\pm 5.47}$ & \hphantom{0}$9.89 {\pm 5.24}$ & $10.90 {\pm 4.65}$ & $15.00{\pm 0.00}$ & $14.88 {\pm 0.65}$ & $14.92 {\pm 0.58}$ \\

SadTalker~\cite{zhang2023sadtalker} & \hphantom{0}$6.11 {\pm 5.74}$ & \hphantom{0}$7.59 {\pm 5.84}$ & \hphantom{0}$9.52 {\pm 4.99}$ & $12.68 {\pm 3.64}$ & $11.26 {\pm 4.24}$ & $10.01 {\pm 4.62}$ \\

\midrule

\method{} (Ours) &  $\textbf{4.55}\pm \textbf{4.91}$ & $\textbf{4.40} \pm \textbf{5.30}$ & $\textbf{7.04} \pm \textbf{5.70}$ & $\textbf{9.01}\pm \textbf{5.45}$ & $\textbf{7.86} \pm \textbf{5.47}$ & $\textbf{6.53} \pm \textbf{5.20}$\\

\bottomrule
\end{tabular}
}
\end{center}
\end{table*}

\paragraph{Results.}

Audio-visual correlation scores provided by SyncNet show strong results for \method{}, significantly ahead of all other methods regarding either the absolute offset or the confidence.
One exception is the confidence score of Wav2Lip, which is biased by the use of the SyncNet score as a loss function.
The LMD gives a complementary picture of the audio-visual alignment, with less variance between models partly due to a lower influence of outputs' visual attributes.
\method{} performs very well regarding this metric, being only surpassed by IP\_LAP on VoxCeleb2 and by SadTalker on LRS2.
Although the novelty of the proposed approach does not lie in the improvement of the fine-scale AV correlation \textit{per se}, the results from Table~\ref{tab:visual_av_vox} show that the use of a pyramid of syncers in the multi-scale loss function has a positive effect on the AV synchrony at the finest time scale.

\subsection{Multi-scale Correlation between Speech and Rigid Head Motion}
\label{rigid_av_correl}

\paragraph{Protocol.}

In this section we explore a research direction generally overlooked in the talking head generation literature.\footnotetext{In all tables, bold indicates best result, underline second best.}
Although several previous works include a mechanism to learn meaningful head motion, none of them attempt to measure the relevance of the produced rigid dynamics.
To tackle this issue, it is once again useful to consider a low-dimensional motion representation: audio-visual input pyramids can be efficiently built in a similar fashion to that of Section~\ref{sec:method}, on a set of points restricted to rigid facial parts.
Although no such subset was used to train our model, it is crucial for a fair comparison with other methods that the multi-scale syncers should not be trained on the same facial keypoints as the ones used to compute the loss.
To this end, we rely instead on the 31 facial landmarks of the eyes and nose extracted from the Face Alignment method~\cite{bulat2017far}, discarding lips and jaws altogether, and further use a triplet loss for the training instead of the cross-entropy loss described in Section~\ref{sec:method}.
Three syncer networks operating on three different time scales were thus trained to quantify the correlation between speech and rigid head motion on both VoxCeleb2~(II) and HDTF, two datasets that preserve motion dynamics.
To prevent unwanted correlations from interfering, \eg between voice pitch and facial structure, contrastive loss pairs are mined from the same sequences.
In addition, all test identities are unseen during training.
Results, reported in Table~\ref{tab:head_av_vox}, are measured on sequences of 80 frames.
For HDTF, a split of 1058 80-frame test clips were extracted from 51 long duration samples chosen randomly for testing.
The AV synchrony is evaluated using the absolute value of the audio-visual offset provided by the newly trained syncer pyramid at three different scales, corresponding to audio-visual chunks of 200~ms, 400~ms and 800~ms.
For 25 fps, this means that an offset of 1 at the finest resolution corresponds to a misalignment of 40~ms between modalities, while this rises to 160~ms at the coarsest time scale.

\paragraph{Results.}
The first important finding is that it is possible to train neural networks to measure the temporal syncing between speech and rigid motion on three different time scales on both datasets.
The evidence of that is the significant difference (\textit{p}-value $ < 0.05$ in all cases) in AV offsets calculated by the syncers between ground truth and randomly sampled audio-visual pairs.
This supports the existence of a correlation between speech and head motion, albeit fainter with HDTF which features videos of codified political addresses.
The second finding is that \method{} consistently performs the best in terms of offset over all other methods, often with a large margin, for all scales of both VoxCeleb2 and HDTF.
This suggests that the proposed method succeeds in producing head motion that are time-aligned with the audio input over various time scales and datasets, which, in average, cannot be said for any other method.

\begin{table}[t]
\caption{Visual quality comparison on VoxCeleb (II) on sequences of 120 frames}
\label{tab:fid}
\begin{center}
\begin{tabular}{l | c }
\toprule
Method &  $\text{FID} \downarrow$ \\
\midrule
MakeItTalk~\cite{zhou2020makelttalk}&  23.6 \\
Wav2Lip~\cite{prajwal2020lip} & 21.9 \\
Audio2Head~\cite{Wang2021Audio2HeadAO} & 28.6 \\
Wang~\etal~\cite{wang2022one} & 19.5 \\
SadTalker~\cite{zhang2023sadtalker} & \textbf{14.1} \\
\midrule
\method{} (ours) & \underline{18.7} \\
\bottomrule
\end{tabular}
\end{center}
\end{table}

\begin{figure*}
  \centering
  \includegraphics[width=1.0\linewidth]{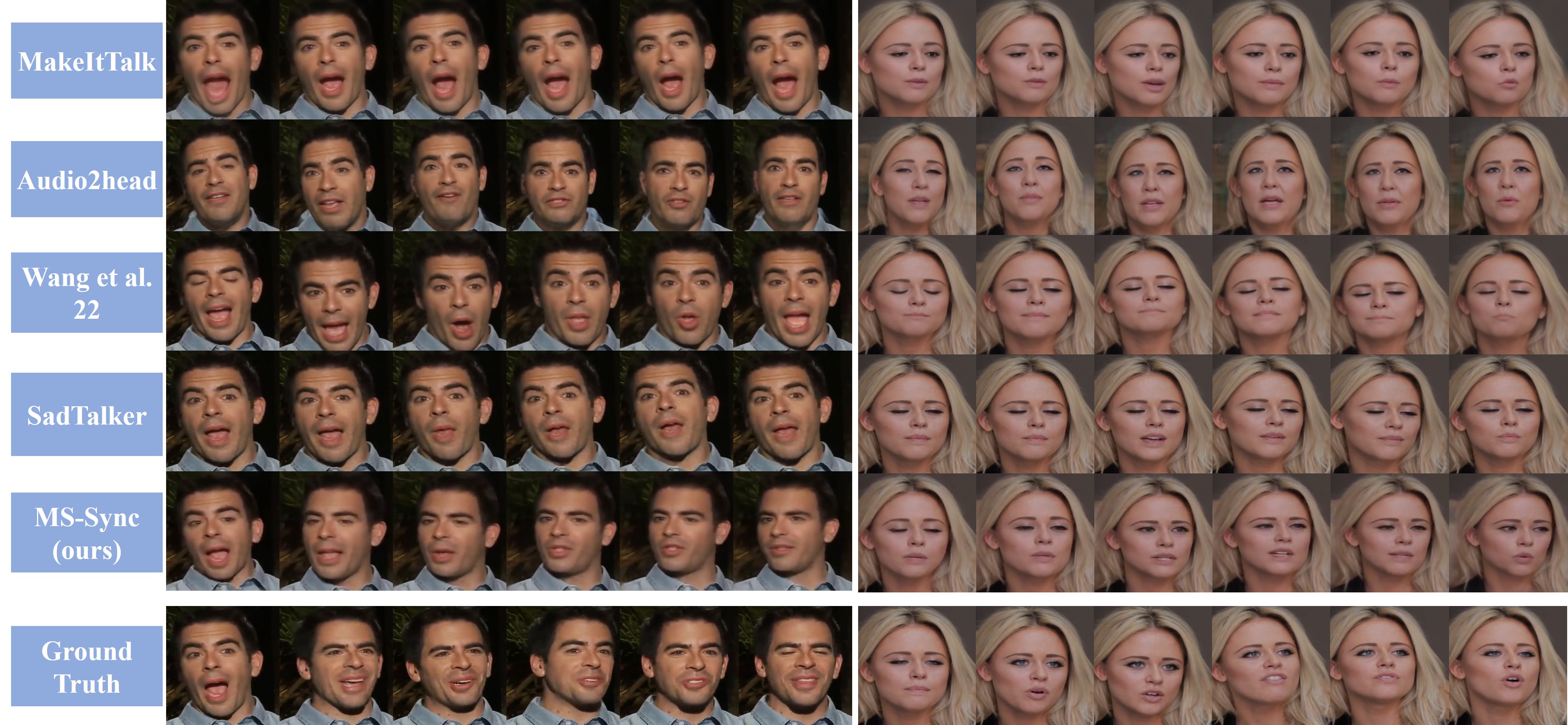}
  \caption{Comparison between different one-shot talking head generation methods with explicit head motion treatment on hard samples: left with a widely open initial mouth, right with closed eyes. Consecutive frames are 600ms apart. \method{} can deal with these samples while producing rigid head motion synchronized with the speech. 
  }
  \label{fig:compa1}
\end{figure*}

\begin{figure*}[tb]
  \centering
  \includegraphics[width=1.0\linewidth]{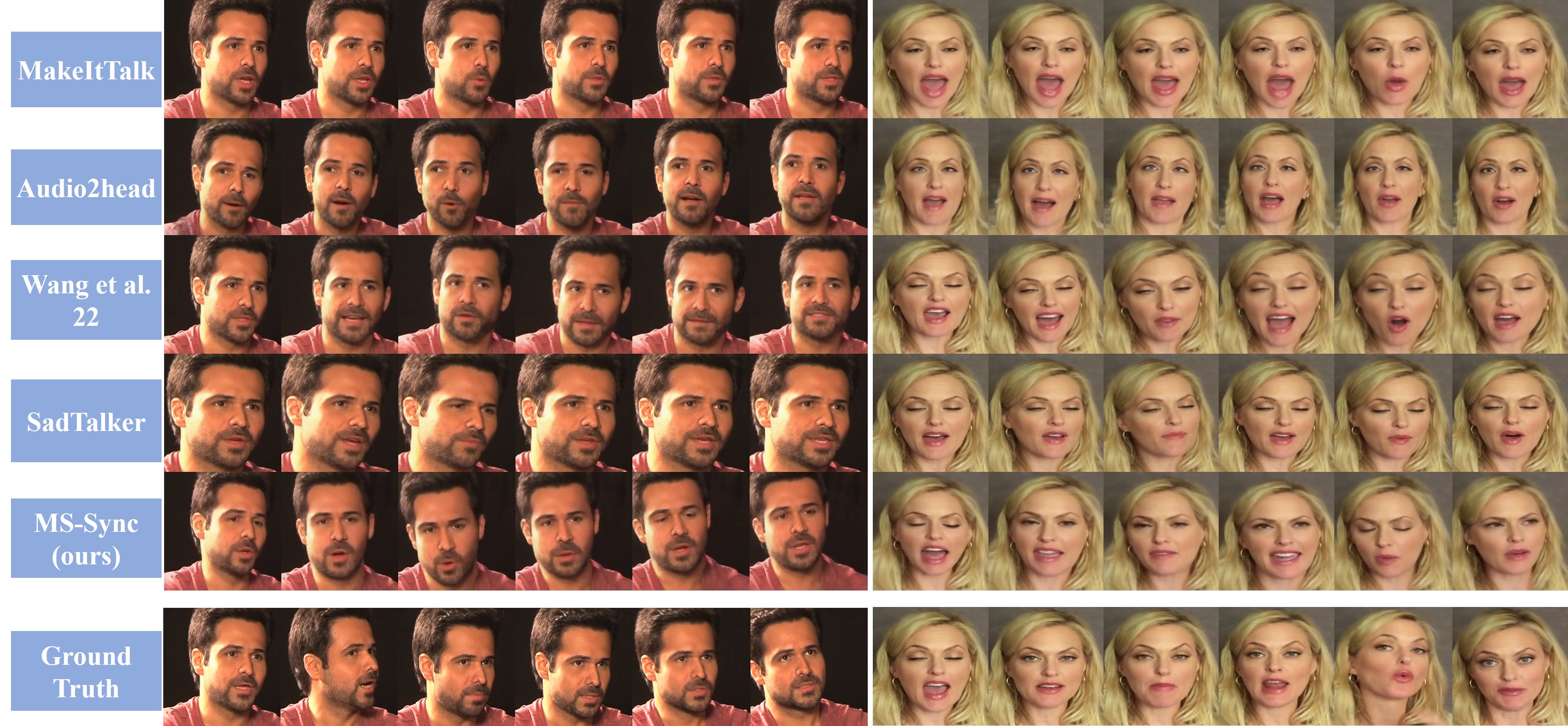}
  \caption{Comparison between different one-shot talking head generation methods with explicit head motion treatment. Consecutive frames are 600ms apart.
  }
  \label{fig:compa2}
\end{figure*}

\subsection{Visual Quality}

\paragraph{Quantitative Results.}
The Fréchet Inception Distance (FID~\cite{heusel2017gans}) of the output sequences is measured and presented it in~\cref{tab:fid}.
We found SSIM and PSNR, which are usually used as complementary metrics of proximity to target images, to correlate poorly with visual sharpness in this free head motion generation setting and therefore do not include them here. 
SadTalker, which produces very sharp outputs based on the 3DMM model, obtains the best results.
However, although using the same reenactment model as that of Audio2Head~\cite{Wang2021Audio2HeadAO} and Wang~\etal~\cite{wang2022one}, \method{} obtains better FID scores, which likely expresses the higher diversity of head pose it produces, better matching the original data distribution.
Interestingly, these results computed on output sequences of 120 frames also highlight the ability of the proposed autoregressive generator to produce accurate keypoints dynamics over extended duration, as training was done on shorter sequences of 40 frames.

\paragraph{Qualitative Comparison.}

Finally qualitative results on VoxCeleb2 are represented in~\cref{fig:compa1} and~\cref{fig:compa2}, along with other one-shot reenactment methods that explicitly handle head motion.
These comprise difficult samples, such as an open mouth or closed eyes in the initial frame.
Contrary to other methods, \method{} deals successfully with all these samples.
MakeItTalk~\cite{zhou2020makelttalk} produces overall good quality results but with almost no head motion.
Audio2head~\cite{Wang2021Audio2HeadAO} struggles to preserve the input identity, a limitation that is overcome in Wang~\etal~\cite{wang2022one} although their strategy of learning on a single identity results in a lack of naturalness and audio correlation.
Finally, SadTalker~\cite{zhang2023sadtalker} produces very sharp video outputs with high quality lip-syncing but fails to correlate output head motion with speech.

\subsection{Ablation Study}

\begin{table}[t]
\caption{Effects of adversarial and reconstruction loss functions on AV correlation and visual quality.}
\label{tab:abla_adv_sup}
\begin{tabular}{c c | c | c }
\toprule
$(\lambda_{\text{rec}}, \lambda_{\text{adv}})$ & $|\text{AV-Off}| \downarrow$ & $\text{AV-Conf} \uparrow$ & $\text{FID} \downarrow$\\

\midrule



$(1, 0)$ & $\underline{0.84} $ & $5.09 $ & 29.6 \\ 

$(1, 0.01)$ & $\textbf{0.75} $ & \textbf{6.40} & 19.7 \\ 

$(1, 1)$ & $1.26 $ & $5.61 $ & \underline{19.0} \\ 

$(0.01, 1)$ & $1.41 $ & $5.52 $ & 20.1 \\ 
\midrule
Full Model $(1, 0.1)$ & $1.06 $ & $\underline{5.75} $ & \textbf{18.7} \\
\bottomrule
\end{tabular}
\end{table}

\begin{figure}
\includegraphics[width=0.9\linewidth]{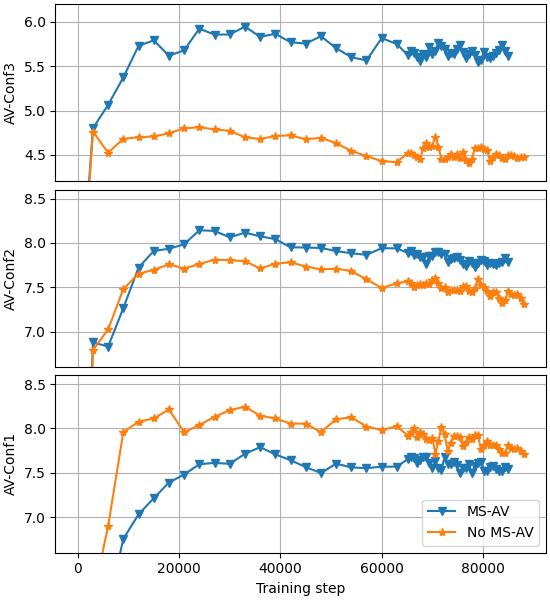}
\centering
\caption{Evolution of multi-scale audio-visual confidence over training, measured on VoxCeleb2 (II) validation set. Bottom is the finest scale, top the coarsest.}
\label{fig:abla_ms}
\end{figure}

\paragraph{Weighting Factors of the Loss Function.}
In this first ablation we explore the effects of the adversarial and the reconstruction terms on the visual quality and the audio-visual synchrony through their respective scaling factors $\lambda_{\text{adv}}$ and $\lambda_{\text{rec}}$ (see \cref{tab:abla_adv_sup}).
Results show that a small, but non-zero, $\lambda_{\text{adv}}$ provides results that are better synced with the input audio, but at the cost of a degraded visual quality.
This is also the case when the scale of the reconstruction is reduced with respect to the adversarial term.
The chosen hyperparameters overall provide the best trade-off, yielding in particular the lowest FID score.

\paragraph{Effects of the Multi-scale AV Loss.}

In \cref{fig:abla_ms} we represent the training evolution of the validation AV confidence over three pyramid layers, and evaluate the effects of removing the multi-scale audio-visual loss, thereby enforcing the synchrony at the finest time scale only (bottom panel in the figure).
One can see that doing so indeed gives improvements in the correlation at the finest time scale, but at the same time the confidence degrades for all coarser time scales.
Finally, using the multi-scale AV loss provides significantly stronger long-range audio-visual synchrony results, achieving the objectives for which it was designed.

\section{Discussion and Limitations}

Experiments show that our framework is an efficient way to explicitly handle the correlation between speech and facial dynamics, including both lips and head motion.
But it also has the advantage of being readily generalizable: the multi-scale AV synchrony loss can fuse in any talking head architecture that relies on an implicit or explicit low-dimensional representation of motion, as long as the use of Gaussian smoothing on this representation makes sense.

\begin{figure}[tb]
  \centering
  \includegraphics[width=0.8\linewidth]{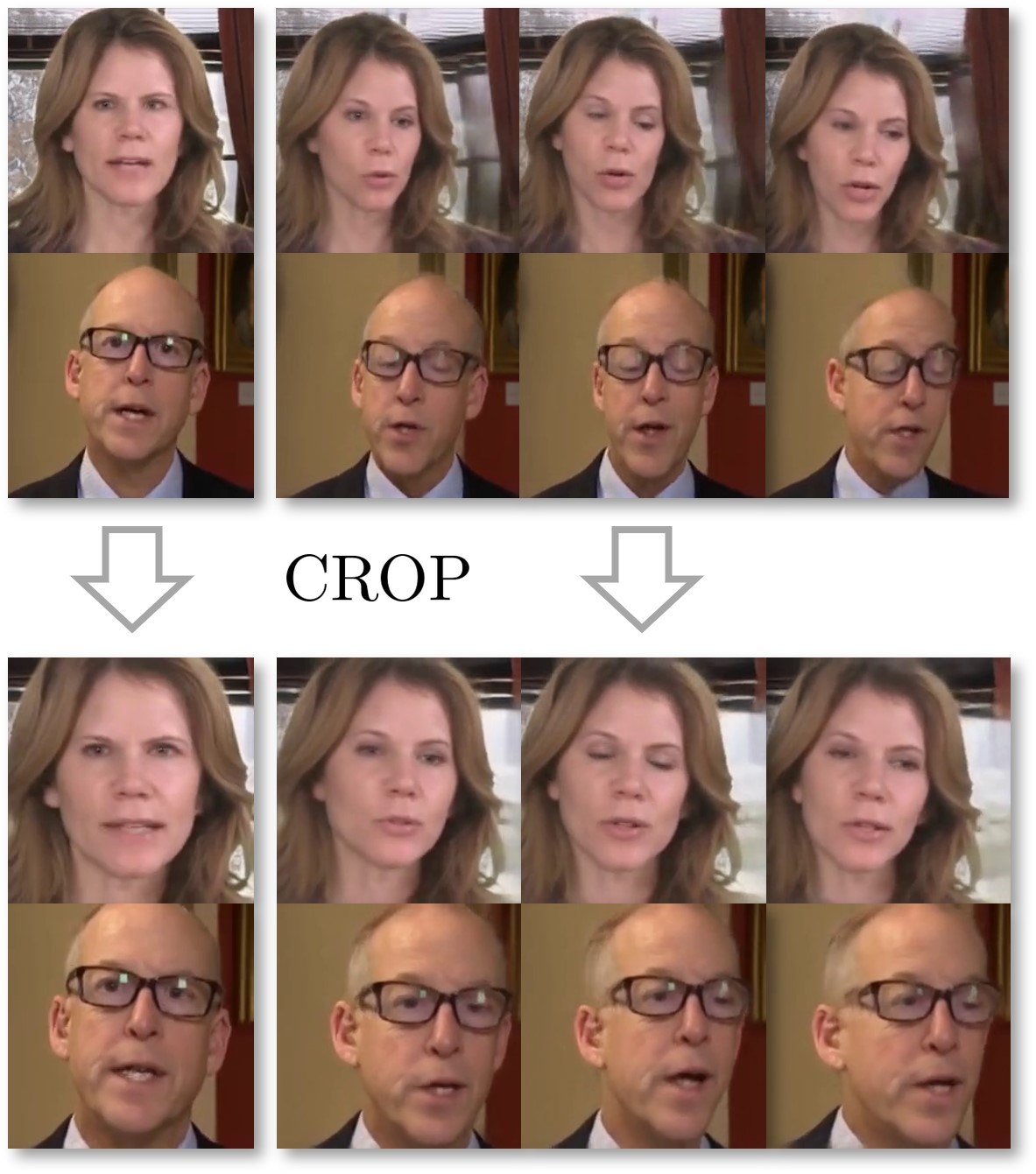}
  \caption{Examples of improvements obtained from cropping the source imagse on two failure cases from HDTF test set.
  }
  \label{fig:hdtf}
\end{figure}

The \method{} model also comes with a number of limitations.
As all continuous autoregressive models, it is prone to error accumulation which limits the output sequence length, although we found the results to remain visually sharp past 120 frames (or $\sim$ 5s) of video duration, which is three times the training sequence length.
Fig. ~\ref{fig:hdtf} provides examples of such failure cases, although in this case a tighter cropping can partially alleviate the visual defaults.
We also found that jitter could be a problem especially on internet images, which, together with the previous observation, could be hinting for a possible sensitivity to out-of-distribution samples.
We hypothesize that propagating visual loss gradients through the reenactment model could alleviate this issue, at the cost of additional computation.

\section{Conclusion}

The approach proposed in this research work is the first attempt to learn and model audio-visual correlations at multiple scales for talking head generation.
For this we use a pyramid of syncer models, trained on hierarchical representations of input audio and head dynamics, which are later used in a multi-scale AV synchrony loss for the training of the generative model.
Experiments showed that the devised multi-scale generative network succeeds in producing realistic head and lip motion output over various time scales.
The very encouraging results of \method{} let us foresee numerous applications of similar approaches on other audio-visual generation tasks, for instance to enforce the consistency of an apparent emotional state which will typically evolve on much longer time scales than the ones considered here.

\bibliographystyle{ieee_fullname}
\bibliography{main}
\end{document}